\documentclass[%
 reprint,
 amsmath,amssymb,
 aip,
 chaos,
]{revtex4-2}

\usepackage{graphicx}
\usepackage{dcolumn}
\usepackage{bm}
\usepackage{color}

\newcommand{\M}{\mathbf{M}}
\newcommand{\W}{\mathbf{W}}
\newcommand{\R}{\mathbf{R}}
\newcommand{\K}{\mathbf{K}}
\newcommand{\OOmega}{\mathbf{\Omega}}

\begin{document}


\title{Reservoir Computing with a Single Oscillating Gas Bubble: Emphasizing the Chaotic Regime}
\author{Hend Abdel-Ghani}
\email{habdelghani@csu.edu.au}
\author{A.~H.~Abbas}
\author{Ivan S.~Maksymov}
\affiliation{Artificial Intelligence and Cyber Futures Institute, Charles Sturt University, Bathurst, NSW 2795, Australia\looseness=-1}

\begin{abstract}
The rising computational and energy demands of artificial intelligence systems urge the exploration of alternative software and hardware solutions that exploit physical effects for computation. According to machine learning theory, a neural network-based computational system must exhibit nonlinearity to effectively model complex patterns and relationships. This requirement has driven extensive research into various nonlinear physical systems to enhance the performance of neural networks. In this paper, we propose and theoretically validate a reservoir computing system based on a single bubble trapped within a bulk of liquid. By applying an external acoustic pressure wave to both encode input information and excite the complex nonlinear dynamics, we showcase the ability of this single-bubble reservoir computing system to forecast complex benchmarking time series and undertake classification tasks with high accuracy. Specifically, we demonstrate that a chaotic physical regime of bubble oscillation proves to be the most effective for this kind of computations.

\end{abstract}

\maketitle


\section{\label{sec:level1}Introduction}
The 2024 Nobel Prize in Physics, awarded to John J.~Hopfield and Geoffrey E.~Hinton for their foundational contributions to artificial neural networks, demonstrates the importance of physics in advancing AI. In modern machine learning (ML) and artificial intelligence (AI), physics plays an increasingly vital role, particularly through physics-informed neural networks \cite{Rai19}, which, for example, are widely applied in areas such as fluid mechanics \cite{Doa21, Pfe22}, quantum neural networks \cite{Per24, Cho24} and neuromorphic computing (NC) systems \cite{Luk09, Tan19, Nak21, Mar20_1, Cuc22}. 
Many novel NC systems have been proposed, employing physical principles from diverse fields such as nanomagnetism \cite{Fur18, Wat20, All23}, quantum mechanics \cite{Mar20_2, Bra22, Gov21, Abb24}, fluid mechanics \cite{Got21, Abb24_review, Mak23_EPL, Mar23, Mat22}, soft matter \cite{Nak13, Nak18}, photonics \cite{Che20_review, Nak21}, electronics \cite{Pen18, Sun21, Cao22, Ike23, Wan23}, acoustics \cite{Mak21_ESN, Hen22} and human-made objects \cite{Lym20, Per24, Abb24_review}.

In this paper, we theoretically demonstrate that a single gas bubble trapped within a bulk of liquid can function as a reservoir computing (RC) system---a type of NC system that employs the nonlinear dynamics of physical systems for unconventional computation \cite{Tan19, Nak21, Cuc22}. Unlike traditional RC models, which are governed by mathematical principles and ML frameworks relying on differential equations where output variations are not directly proportional to input changes, RC systems based on fluid-mechanical physical objects \cite{Got21, Abb24_review, Mak23_EPL, Mar23, Mat22} offer a more energy-efficient and computationally efficient alternative.

Energy efficiency is a critical consideration in resolving the growing challenge of rapidly increasing energy consumption by AI systems \cite{Ver23, Abb24_review}. In the case of a single bubble located far from boundaries---as in our study---nonlinear oscillations of the bubble can lead to intriguing phenomena such as chemical reactions and the conversion of sound into light (sonoluminescence) \cite{Akh97, Lau10, Tan11, Mak22_SL}. While our work may not necessarily exploit all these effects to their full extent, by using the inherent nonlinearity and energy efficiency of bubble dynamics, our approach offers a promising avenue for developing sustainable and scalable computing solutions. 

Many RC systems draw inspiration from the functioning of a biological brain that operates through vast, intricate networks of neural connections \cite{Mck94, Kor03}. Like the brain, an RC system is a dynamical systems meaning it evolves over time and exhibit complex, nonlinear and sometimes chaotic behavior \cite{Maa02, Jae04, Kra23}. Subsequently, this work specifically focuses on a bubble oscillation regime that induces chaotic dynamical behavior, optimizing the computational capabilities of the RC system.
\begin{figure*}
    \centering
    \includegraphics[width=0.9\linewidth]{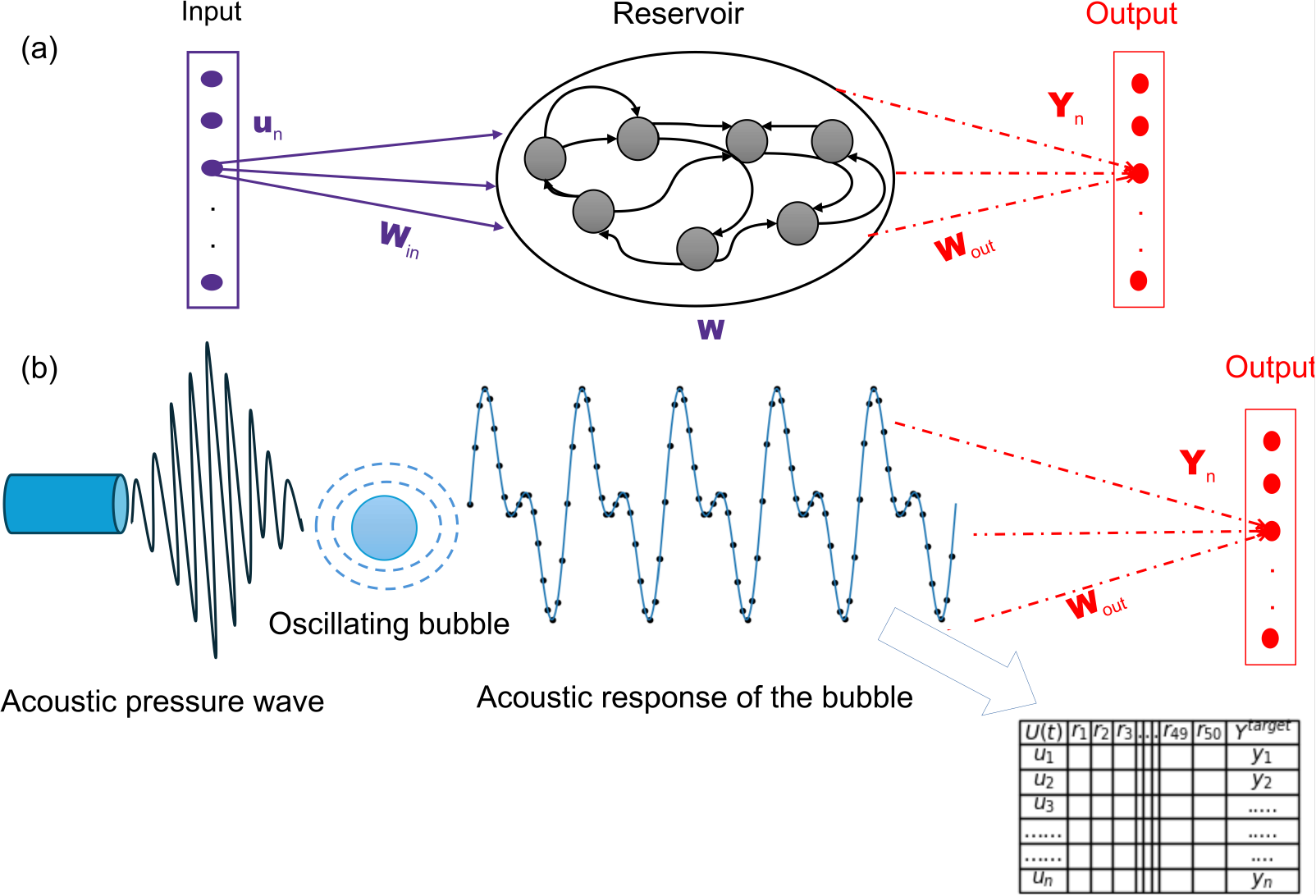}
    \caption{(a)~Schematic illustration of a traditional RC system. The reservoir consists of a network of interconnected artificial neurons that generate a vector of neural activations $\mathbf{x}_n$ from a dataset of input values $\mathbf{u}_n$. Only the linear readout is trained to produce the output $\mathbf{y}_n$. (b)~Bubble-based RC system. The input data are encoded in the peak amplitude of the acoustic pressure waves. Neural activations are extracted by sampling the acoustic response of the oscillating bubble, as detailed in the main text and summarized in the inset table. The training and exploitation procedures for the physical RC system mirror those of the traditional algorithmic RC system.}
    \label{fig4}
\end{figure*}

A previous theoretical study \cite{Mak21_ESN} explored the nonlinear dynamics of an acoustically excited bubble cluster, treating each bubble as a node within a virtual network of interconnected oscillators, analogous to neural network architectures typically employed in RC systems \cite{Mak23_review}. Building on an earlier investigation of bubble clusters with similar equilibrium radii \cite{Tony21}, it was proposed that, for a short yet sufficient duration to conduct measurements, the cluster would maintain stability, making practical implementation feasible. Experimental follow-up research (unpublished) confirmed the feasibility of this approach in principle. However, it was found that more reproducible results could be achieved using only a few interacting bubbles rather than a larger cluster. These findings, combined with a theoretical demonstration of substantial computational memory possessed by an oscillating bubble \cite{Mak23_unconv}, further motivate the present study on a single-bubble RC system, highlighting its potential for practical applications using relatively simple equipment well-suited for an in-depth analysis of single-bubble properties \cite{Gai92, Lau10, Mak21}.

The remainder of this paper is organized as follows. In Sect.~\ref{sec:Theory}, we present the conventional RC algorithm, providing key reference information to aid in understanding both the computational and physical aspects of the single-bubble RC system. Additionally, we introduce the mathematical model of a single oscillating bubble and discuss the relevant physical operating regime for this study. In Sect.~\ref{sec:BRC}, we present the bubble-based RC algorithm and justify the choice of benchmarking tasks adopted in this work. The main results and their analysis are presented in Sect.~\ref{sec:Results}, followed by the conclusions and recommendations for future work.

\section{\label{sec:Theory}Theory}
\subsection{Traditional Reservoir Computing}
Figure~\ref{fig4}(a) illustrates the structure of the reference traditional RC system, where the states of the randomly connected nodes evolve over time under nonlinear dynamics described by the following differential equation \cite{Luk09, Luk12}
\begin{equation}
  {\bf x}_{n} = (1-\alpha){\bf x}_{n-1} + 
  \alpha\tanh({\bf W}^{in}{\bf u}_{n} + {\bf W}{\bf x}_{n-1})\,,
  \label{eq:1}
\end{equation}
where $n$ is the sequential index of the discrete, equally spaced time steps $t_n$, ${\bf u}_n$ is a vector of $N_u$ input values, and ${\bf x}_n$ is a vector of $N_x$ neural activations. The element-wise operator $\tanh(\cdot)$ serves as the activation function. The input weight matrix ${\bf W}^{in}$, randomly generated with dimensions $N_x \times N_u$, represents the input weights, while the recurrent weight matrix ${\bf W}$, of size $N_x \times N_x$, defines the interconnections among the network nodes. The parameter $\alpha \in (0, 1]$ is the leaking rate that regulates the temporal dynamics of the system.

To compute the output weights ${\bf W}^{out}$, we solve the linear equation ${\bf Y}^{target} = {\bf W}^{out}{\bf X}$, where the state matrix ${\bf X}$ is constructed from the neural activations ${\bf x}_n$ and the target matrix ${\bf Y}^{target}$ contains the corresponding target outputs ${\bf y}_n^{target}$ for each time step $t_n$. The output weights ${\bf W}^{out}$ are computed as
\begin{equation}
{\bf W}^{out} = {\bf Y}^{target} {\bf X^\top} ({\bf X}{\bf X^\top} + \beta {\bf I})^{-1}\,,\label{eq:Wout}
\end{equation}
where ${\bf I}$ is the identity matrix, $\beta$ is a regularization coefficient and ${\bf X^\top}$ denotes the transpose of ${\bf X}$ \cite{Luk12}. Once the output weights ${\bf W}^{out}$ have been determined, they are used to compute the output vector ${\bf y}_n$ for new input data ${\bf u}_n$, according to the equation \cite{Luk12}
\begin{equation}
{\bf y}_n = {\bf W}^{out}[1; {\bf u}_n; {\bf x}_n]\,.
\end{equation}

Several additional requirements must be met to create an efficient RC system \cite{Jae01, Luk12}. Firstly, the reservoir weights ${\bf W}$ have to be scaled to ensure the echo state property, which guarantees that the neural states of the reservoir are stable and dependent on the input history \cite{Luk09, Luk12}. This is typically achieved by setting the spectral radius $\rho$ of ${\bf W}$ to be less than 1:~$\rho = \max |\lambda|$, where $\lambda$ are the eigenvalues of ${\bf W}$. In this context we note that nonlinear dynamics of a bubble inherently satisfy the echo state property \cite{Mak21_ESN, Mak23_review}, enabling the RC system proposed in this paper to function as an efficient computational framework.

Furthermore, software implementing the traditional RC algorithm must ensure that different inputs are mapped to distinct reservoir states, while similar states map to the same output, thereby enhancing robustness against minor variations. These properties are typically achieved using a large network of reservoir nodes \cite{Luk12} (typically more than 1000). However, as will be demonstrated later in this paper, a physical implementation of the RC system relaxes the requirement for a large number of nodes, making the physical system more computationally and energy efficient.

\subsection{\label{sec:level3}Nonlinear dynamics of a single bubble} 
\subsubsection{Keller-Miksis Equation}
The study of the dynamics of a single gas bubble has a rich history rooted in fluid mechanics and acoustics. Pioneering work by Lord Rayleigh laid the foundation for understanding bubble collapse \cite{Ray17}, where he derived an equation to describe the radial motion of a spherical bubble in an incompressible fluid. This was later extended by Plesset and others to include effects such as surface tension \cite{Col48, Ple49}, viscosity and compressibility, resulting in the well-known Rayleigh-Plesset equation and its modifications, including the Keller-Miksis equation \cite{Kel80, Lau10}. Additionally, researchers like Minnaert explored the acoustic oscillations of bubbles, leading to the formulation of the Minnaert frequency, which characterizes the natural resonance of a gas bubble in a liquid \cite{Min33}.

To date, the nonlinear dynamical properties of oscillating bubbles have remained the subject of substantial theoretical and experimental research \cite{Pro74, Pro77, Kel80, Pro84, Par90, Lau99, Lau10, Lei12}, including phenomena such as cavitation \cite{Bre95}, sonoluminescence \cite{Akh97, Lau10, Tan11, Mak22_SL}, bubble collapse-induced shock waves \cite{Met97} and the translational motion of bubbles \cite{Doi01, Doi02}. These studies have profound implications across various fields, ranging from medical applications like ultrasound imaging and drug delivery \cite{Sus12, Soj21} to industrial processes such as sonochemistry \cite{Tan11, Soj12, Hag19}.

Numerical models of bubbles trapped in a bulk of liquid and forced by an external acoustic pressure field are well-known and thoroughly documented \cite{Ple49, Kel80, Par90, Bre95, Sus12, Mak22_Bio, Mak22_combs}. In the following, we consider a single bubble with an equilibrium radius $R_0$ suspended in an incompressible, viscous liquid. When exposed to an external, time-varying pressure field $P_\infty(t)$, the instantaneous radius of the bubble $R(t)$ undergoes an oscillatory motion, with dynamics governed by the Keller-Miksis equation \cite{Kel80, Sus12}
\begin{multline}
\left(1 - \frac{\dot{R}}{c}\right)R\ddot{R} + \frac{\dot{R}^2}{2} \left(3 - \frac{\dot{R}}{c} \right) = \\
\frac{1}{\rho} \left(1 + \frac{\dot{R}}{c}+ \frac{R}{c} \frac{d}{dt} \right) 
\left[ P(R, \dot{R}) - P_{\infty}(t)\right]\,, \label{eq:KMP}
\end{multline}
where $\dot{R}$ and $\ddot{R}$ are the first and second time derivatives of $R(t)$, $c$ is the speed of sound in the liquid and $\rho$ is the fluid density. The internal pressure of the bubble is defined as
\begin{equation}
P(R, \dot{R}) = \left( P_0 - P_v + \frac{2\sigma}{R_0} \right)\left(\frac{R_0}{R}\right)^{3\kappa} - \frac{2\sigma}{R}-\frac{4\mu}{R} \dot{R}\,.
\end{equation}
The external pressure is 
\begin{equation}
P_{\infty}(t) = P_0 - P_v + P_a \sin(\omega t)\,,
\end{equation}
where $P_0$ denotes ambient pressure, $P_v$ is the vapor pressure inside the bubble, $P_a$ represents the amplitude of the acoustic pressure and $\omega = 2\pi f_a$ is the angular frequency of the driving acoustic pressure wave. The initial conditions are given by $R(0) = R_0+\tilde{R}_0, \dot{R}(0)=V$. The remaining model parameters are the dynamic viscosity of the liquid $\mu$, the polytropic exponent $\kappa$ for the gas inside the bubble and the surface tension $\sigma$ at the gas-liquid interface (Table~\ref{tab:simulation_properties}).

The natural frequency of the bubble is \cite{Mak21}
\begin{equation}
f_{nat} = \frac{1}{2\pi\sqrt{\rho}R_0}\sqrt{3\kappa\left(P_0-P_v+\frac{2\sigma}{R_0}\right)-\frac{2\sigma}{R_0}-\frac{4\mu^2}{\rho R_0^2}}\,.
\end{equation}
This expression can be recast as
\begin{equation}
f_{nat} \approx f_M \left(1+\frac{(3\kappa-1)\sigma}{3\kappa R_0(P_0-P_v)}-\frac{2\mu^2}{3\kappa \rho R_0^2(P_0-P_v)}\right)\,,
\end{equation}
where $f_M = \dfrac{\sqrt{3\kappa(P_0-P_v)}}{2\pi R_0 \sqrt{\rho}}$ is the well-known Minnaert frequency \cite{Min33}.

We nondimensionalize Eq.~(\ref{eq:KMP}) to reduce the number of governing parameters \cite{Sus12}. This is done by using the equilibrium radius $R_0$ and the inverse of the ultrasound frequency $\omega^{-1}$ as characteristic length and time scales, respectively. We define the nondimensional bubble radius $r$ and time $\tau$ as $r = \dfrac{R(t)}{R_0}$ and $\tau = \omega t$, respectively. Thus, we obtain
\begin{gather}
\ddot{r} \left[(1 - \OOmega \dot{r})r + \OOmega \R\right] 
= \left(\OOmega \dot{r} - 3 \right) \frac{\dot{r}^2}{2} - \frac{\W + \R \dot{r}}{r} \quad \notag \\
+ (\M + \W) \frac{\left[1 + (1 - \K) \OOmega \dot{r}\right]} {r^{\K}} \notag \\
- (1 + \OOmega \dot{r})(\M + \M_e \sin \tau) - \M_e \OOmega r \cos \tau\,, \label{eq:non_dim}
\end{gather}
where the nondimensional parameters are defined as
\begin{align}
\OOmega &= \frac{\omega R_0}{c}, \quad \R = \frac{4\mu}{\rho \omega R_0^2}, \quad \W = \frac{2\sigma}{\rho \omega^2 R_0^3}, \notag\\
\M &= \frac{P_0 - P_v}{\rho \omega^2 R_0^2}, \quad \M_e = \frac{P_a}{\rho \omega^2 R_0^2}, \quad \K=3\kappa.
\end{align}

Each of the nondimensional groups listed above has a straightforward physical meaning. The parameter $\OOmega$, which is the ratio of the equilibrium bubble radius to the acoustic wavelength, characterizes the bubble size. Parameters $\R$ and $\W$ characterize the viscous dissipation and surface tension effects, respectively, and can be treated as inverse Reynolds and Weber numbers. The parameter $\M$ represents the elastic properties of the gas and its compressibility, while $\M_e$ measures the external acoustic excitation. Additionally, the nondimensional Minnaert frequency can conveniently be expressed as $\omega_0=\sqrt{\K\M}$ \cite{Mak21}.

These parameters enable us to systematically analyze the behavior of the bubble under different physical conditions, making the study of bubble dynamics in a liquid more general and widely applicable. We numerically solve Eq.~(\ref{eq:non_dim}) using the \texttt{odeint} procedure from the SciPy library of Python 3.0 programming language using the material parameters listed in Table~\ref{tab:simulation_properties}.
\begin{table}[ht]
    \centering
    \caption{Model parameters used this study}
    \begin{tabular}{|l|c|c|}
        \hline
        \textbf{Parameter} & \textbf{Value} & \textbf{Unit} \\
        \hline
        Density of water ($\rho$) & $998$ & kg/m$^3$ \\
        \hline
        Static pressure ($P_{\text{stat}}$) & $100 \times 10^3$ & Pa \\
        \hline
        Vapor pressure ($P_v$) & $2.33 \times 10^3$ & Pa \\
        \hline
        Surface tension ($\sigma$) & $7.25 \times 10^{-2}$ & N/m \\
        \hline
        Gas polytropic exponent ($\kappa$) & 1.4 & - \\
        \hline
        Driving acoustic frequency ($f_a$) & $6.2362 \times 10^6$ & Hz \\
        \hline
        Dynamic viscosity ($\mu$) & $1 \times 10^{-3}$ & kg/(m·s) \\
        \hline
        Equilibrium bubble radius ($R_0$) & $0.8 \times 10^{-6}$ & m \\
        \hline
        Velocity of sound in water ($c$) & $1.50 \times 10^3$ & m/s \\
        \hline
    \end{tabular}
    \label{tab:simulation_properties}
\end{table}

\subsubsection{Physical operating regimes of interest}
\begin{figure}[t]
    \centering
    \includegraphics[width=1\linewidth]{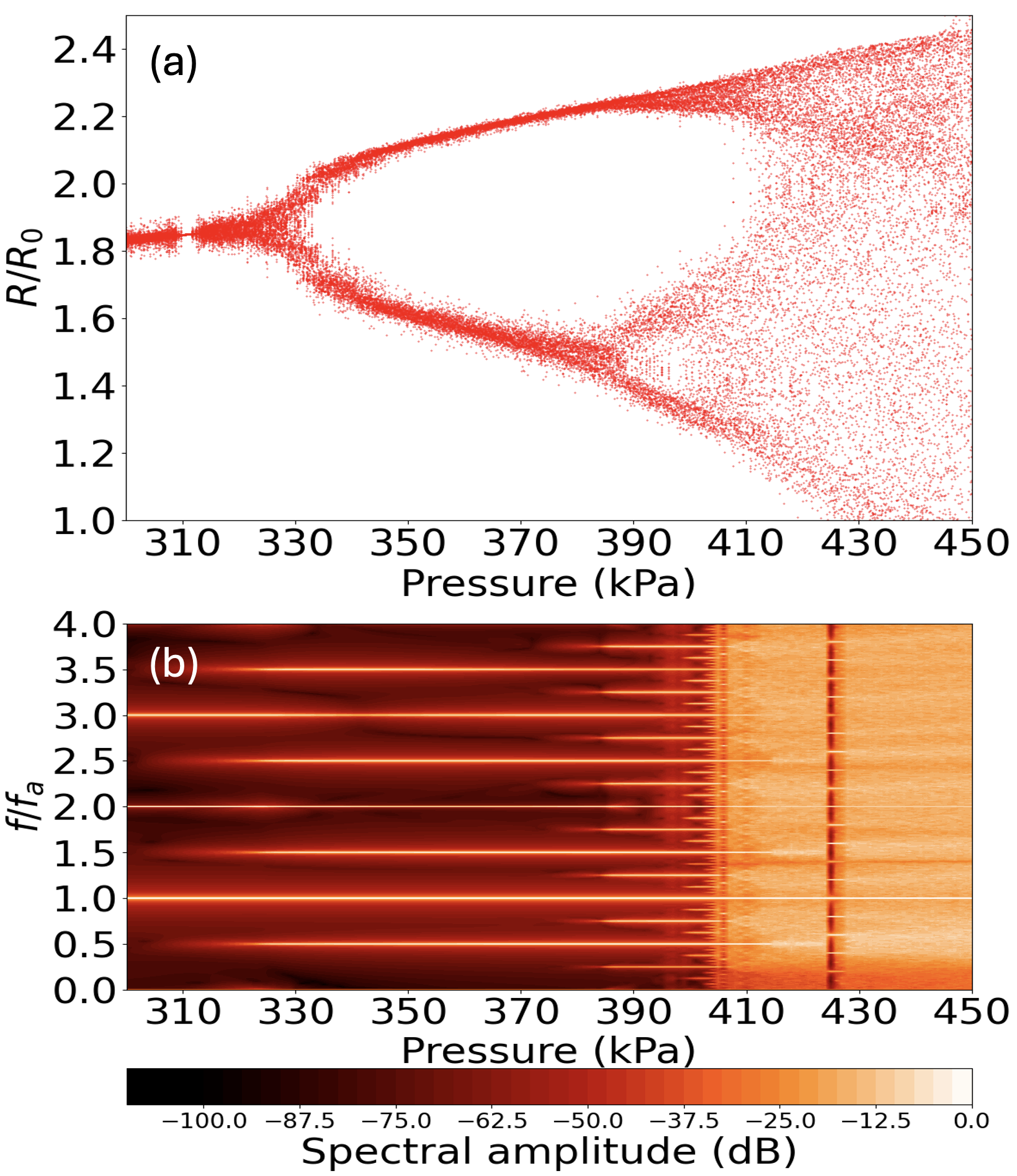}
    \caption{(a)~Theoretical bifurcation curve of a single bubble functioning as an RC system and excited with a sinusoidal acoustic wave with the peak pressure $P_a$. (b)~Spectral representation of the nonlinear dynamical regimes of oscillation.}
    \label{fig11}
\end{figure}

Figure~\ref{fig11}(a) presents the bifurcation diagram \cite{Par90, Lau10, Sus12}, depicting the normalized bubble radius $R/R_0$ as a function of the peak driving acoustic pressure $P_a$, systematically varied from $300$\,kPa to $450$\,kPa in $0.1$\,kPa increments. The equilibrium bubble radius, $R_0 = 0.8$\,$\mu$m, is chosen to reflect typical conditions in both research and industrial applications, ensuring a realistic study context \cite{Sus12}. This diagrams visually depict how the response of the system evolves as $P_a$ is varied at a constant value of the driving acoustic frequency $f_a$ (Table~\ref{tab:simulation_properties}). 

Providing complementary information, Fig.~\ref{fig11}(b) presents the evolution of the acoustic spectrum of the bubble, where the $x$-axis corresponds to the normalized frequency $f/f_a$. The $y$-axis represents the peak pressure $P_a$ of the incident wave but the false color encodes the amplitude (in dB) of the acoustic pressure scattered by the bubble \cite{Mak21}. We can see that at lower pressures, the spectrum shows frequency peaks at $f/f_a = 1, 2, 3$ and so on (the nondimensional Minnaert frequency is $\omega_0=0.8624$). As the peak pressure increases, subharmonic peaks at $f/f_a = \frac{1}{2}$ and their ultraharmonic components at $f/f_a = \frac{3}{2}, \frac{5}{2}$, and so on, become more prominent. We also observe that the emergence of additional peaks in the spectrum correlates with the bifurcation regions identified in the diagram in Fig.~\ref{fig11}(a).

In Fig.~\ref{fig11}(a), we can see that at low values of $P_a$, oscillations exhibit simple periodic behavior, referred to in this paper as single-period oscillations. As the pressure increases to 330\,kPa, the bubble transitions into a more complex, double-periodic oscillation regime. Further increasing the driving pressure to approximately 385\,kPa leads to a quadratic periodic regime. When the pressure reaches approximately 400\,kPa, the system enters a chaotic regime, where the radial oscillations become highly irregular and unpredictable, which is especially seen for $P_a \ge 430$\,kPa.

We also established that the so-called edge-of-the-chaos regime, occurring within the pressure range of $P_a = 400\dots410$\,kPa and marked by the transition from a periodic state to a chaotic one, is particularly promising for applications in RC systems. In this regime, the response of the bubble exhibits a trade-off between order and chaos, enabling it to respond dynamically to inputs without becoming entirely unpredictable or overly sensitive to small perturbations. 

We now discuss the significance of these physical regimes in the context of ML. The specific nonlinear regimes suitable for the operation of both traditional and physical RC systems have been extensively discussed in the literature \cite{Luk09, Tan19, Nak20}. At the same time, it has been shown that RC systems \cite{Maa02, Ber04, Nis22, Abb25}, particularly those based on fluid-mechanical systems \cite{Mak23_review}, can operate using a broader range of nonlinear phenomena.   

In the periodic regime of single-bubble oscillations, the physical system exhibits high stability and predictability, making it resistant to small variations in initial conditions. However, when used as an RC system, this stability reduces flexibility and limits the ability of the system to capture complex patterns, leading to the undesired effect of underfitting.

In contrast, the chaotic regime of the system is highly sensitive to initial conditions, where small differences in input can lead to vastly different outputs. This sensitivity enables the system to model diverse behaviors, enhancing pattern recognition, but it also increases the risk of amplifying noise and overfitting. While this sensitivity improves pattern recognition, it makes the system less reliable when generalizing to new data.

Thus, the choice of operating regime significantly influences the performance of an RC system. A highly stable system may struggle to learn effectively, while an overly chaotic one becomes unreliable. Subsequently, in this paper, the-edge-of-chaos regime is regarded as the optimal regime, as it strikes a crucial balance, enabling the reservoir to model and predict real-world data efficiently.


\section{\label{sec:BRC} Bubble-Based Reservoir Computing}
In this section, we introduce the algorithm for the bubble-based RC system. Since our approach builds on specific computational steps from the traditional RC procedure outlined above, we reference the computational aspects of the traditional system, as summarized in Fig.~\ref{fig4}. Additionally, we explain the rationale behind the choice of benchmarking tasks and discuss the computational operating regimes of the RC system suitable for these tasks.

\subsection{Physico-computational framework}
From a theoretical perspective, implementing a single bubble as a reservoir involves replacing the core update equation of the traditional RC system [Eq.~(\ref{eq:1})] with Eq.~(\ref{eq:non_dim}), which governs the nonlinear dynamics of the bubble. This step is complemented by the proper encoding of input data as the driving signal that forces oscillations of the bubbles, as well as the computation and simultaneous processing of the acoustic pressure waves emitted by the bubble, play a key role in forming the output data [Fig.~\ref{fig4}(b)]. Additionally, as demonstrated in prior work \cite{Mak21_ESN, Mak23_review}, such a replacement enables relaxing certain computational requirements, such as the calculation of the spectral radius. This is because, similar to other physical RC systems \cite{Luk09, Tan19}, the nonlinear dynamics of the bubble inherently satisfies the echo state property, which is central to the concept of reservoir computing.

To implement the aforementioned procedures, we encode the input signal $u(t)$ of interest, with discrete time steps $t_n$ implied, as the peak pressure amplitude of a sinusoidal acoustic wave [Fig.~\ref{fig4}(b)]. The input signal $u(t)$ is first normalized to the interval $[0, 1]$ and then mapped to the peak acoustic pressure amplitude in the range $[P_{\text{min}}, P_{\text{max}}]$, which represents the boundary values of the physical operating regime inferred from the analysis of the bifurcation diagram and Fourier spectra, as discussed above.

The normalization procedure is implemented by computing
\begin{equation}
u_{\text{norm}}(t) = \frac{u(t) - u_{\text{min}}}{u_{\text{max}} - u_{\text{min}}}\,,
\end{equation}
where $u_{\text{max}}$ and $u_{\text{min}}$ are the maximum and minimum values in the dataset of interest, respectively. The so-obtained signal is then used in the transformation that produced a discrete-time function
\begin{equation}
U(t) = P_{\text{min}} + (P_{\text{max}} - P_{\text{min}}) \cdot u_{\text{norm}}(t)\,
\end{equation}
to ensure the proper scaling of the input signal required to match the operational parameters, such that $ u_{\text{norm}}(t) \in [P_{\text{min}}, P_{\text{max}}]$. 

To implement the RC system, independent simulations of bubble dynamics are conducted, with the number of simulations corresponding to the number of discrete points in $U(t)$ [see the inset table in Fig.~\ref{fig4}(b)]. The function $U(t)$ contains the values of the peak pressure of the sinusoidal acoustic wave that drives the oscillations of the bubble: each value of $U(t)$ is used in one simulation.

In each individual computational run of the model that numerically solves Eq.~(\ref{eq:non_dim}), the bubble is subjected to a fixed number of empirically chosen (typically 65) cycles of the sinusoidal signal forcing (the choice of the initial conditions for each run will be discussed below). A temporal signal associated with numerically simulated radial oscillations of the bubble is recorded and transient effects are removed to achieve stable system dynamics. The steady-state portion of each signal is then divided into $N = 5$ segments and then $k = 10$ evenly spaced points are chosen within each segment, thereby producing a total of $k \times N = 50$ discrete data points (these data points are visualized in Fig.~\ref{fig4}(b) as the dot markers superimposed on the representative acoustic response of the bubble). Importantly, when arranged into the vector ${\bf r}_{n} = [r_{1}, r_{2}, \dots , r_{k \times N}]$, the so-prepared data points serve as the neural activations of the reservoir and correspond to the activation states ${\bf x}_{n}$ of the traditional reservoir system [see Eq.~(\ref{eq:1})].

Then, following the training procedure employed in the traditional RC algorithm, we calculate the output weights $\mathbf{W}^{\text{out}}$ between the target point $\mathbf{y}_n^{\text{target}}$ and the corresponding points contained in the vector ${\bf r}_{n}$ using a linear regression procedure. To this end, we solve the linear equation $ \mathbf{Y}^{\text{target}} = \mathbf{W}^{\text{out}} \mathbf{R} $, where the state matrix $\mathbf{R}$ contains all the neural activations ${\bf r}_{n}$ derived from the nonlinear response of the oscillating bubble but the target matrix $\mathbf{Y}^{\text{target}}$ contains the corresponding target outputs $\mathbf{y}_n^{\text{target}}$ for each discrete time step $t_n$. 

The output weights $ \mathbf{W}^{\text{out}} $ are then computed as
\begin{equation}
\mathbf{W}^{\text{out}} = \mathbf{Y}^{\text{target}} \mathbf{R}^\top \left( \mathbf{R} \mathbf{R}^\top + \beta \mathbf{I} \right)^{-1}\,,
\end{equation}
i.e.~by recasting Eq.~(\ref{eq:Wout}) and applying it to the data derived from the numerical model of the oscillating bubble. Continuing drawing analogy with the traditional RC algorithm, once the output weights $\mathbf{W}^{\text{out}}$ have been determined, we use them to compute the predicted output vector $\hat{\mathbf{y}}_n$ for new, i.e.~unseen by the trained RC system, input data $\mathbf{u}_n$ as
\begin{equation}
\hat{\mathbf{y}}_n= \mathbf{W}^{\text{out}} \mathbf{r}_n\,.
\end{equation}

Thus, the goal of the training process is to determine $W_{\text{out}}$ by minimizing a loss function, which is typically achieved by computing the normalized mean squared error between the predicted outputs and the actual target values as
\begin{equation}
 \text{NMSE} = \frac{1}{N} \sum_{i=1}^{N} \left(y_i^{\text{target}} - \hat{y}_i \right)^2\,,
\end{equation}
where $N$ is the number of training samples. It is worth noting that in many test problems, including those discussed below, the target data are obtained by splitting a known time series into two parts: the first part is used for training, and the second part is used for testing. Unless discussed otherwise, during training the system is presented only with the training portion, while the testing portion remains unknown to it. However, the testing portion is known to the human operator, who may use it to compute NMSE to evaluate the accuracy of the model.    

We also discuss the initial physical conditions used in each individual computational run, as outlined in the algorithm above. When the bubble-based RC system operates in a periodic regime, the initial conditions are set once in the very first computational run, ensuring that the system dynamics remains continuous throughout the operation. However, when the RC system operates in a chaotic regime, it is reset to the same initial condition between each input. This ensures that the chaotic dynamics is consistently maintained for each new input signal \cite{Ber04, Abb25}.

\subsection{Benchmarking tasks}
Typically, the accuracy of forecasts made by RC systems has been assessed by evaluating their ability to learn and predict highly nonlinear and chaotic time series of natural, mathematical and synthetic origin. Examples of such benchmarking tasks include the Mackey-Glass time series \cite{Luk12, Mak23_review, Abb25}, the Lorenz \cite{Gau21, Har24} and R{\"o}ssler attractors \cite{Mak21_ESN, Har24}, and the H{\'e}non \cite{Wan23, Sun23} and Ikeda \cite{Mak21_ESN} maps. Each of these tests has its own advantages and limitations, and their selection is often task- and RC-system-specific, relying mostly on heuristics.

In this paper, as the principal benchmarking task we employ the H{\'e}non map, a two-dimensional discrete-time dynamical system known for its chaotic behavior \cite{Hen76}. The recursive equations of this mathematical model are
\begin{equation}
    x_{n+1} = 1 - ax_n^2 + y_n, \quad y_{n+1} = b x_n\,,
\end{equation}
where the typical parameters are $a = 1.4$ and $b = 0.3$.

In the context of RC systems, the H{\'e}non map test offers distinct advantages over the other commonly used benchmarks. Firstly, the H{\'e}non map has a lower dimensionality since it is a discrete-time system governed by only two variables (e.g., the Lorenz and R{\"o}ssler attractors consist of a set of three coupled differential equations \cite{Lor63}). This reduced dimensionality lowers computational costs, making this benchmarking task particularly suitable for efficiently evaluating RC performance in autonomous systems \cite{Abb24_review}.

Secondly, the nonlinear and chaotic dynamics of the H{\'e}non map is captured through a relatively simple recurrence relation, compared with the Lorenz and R{\"o}ssler systems that require numerical integration due to their continuous nature. This feature aligns well with many practical applications of RC systems where data are naturally sampled at discrete time intervals (e.g., sequence prediction, financial time series forecasting and digital signal processing \cite{Wat21}).

Thirdly, the H{\'e}non map exhibits sharp transitions and sudden regime shifts, providing a valuable test for assessing the short-term memory capacity of an RC system \cite{Luk09, Luk12}. This characteristic makes this test task particularly useful for evaluating architectures designed to handle rapid fluctuations and abrupt changes in dynamical systems \cite{Mak23_EPL}. Additionally, a well-defined structure and relatively low-dimensional nature of a H{\'e}non map facilitate interpretability and benchmarking, enabling a human operator on an RC system to isolate and analyze key aspects of RC performance without the added complexity of high-dimensional attractors \cite{Mak23_review}.

Of course, the Lorenz system and the other popular test tasks remain an important benchmark for RC systems, particularly for tasks that require long-term stability in chaotic forecasting or involve continuous system dynamics \cite{Gau21}. Nevertheless, when the focus is on computational efficiency, discrete-time compatibility and the ability to capture sharp nonlinear transitions, which is the case of this present paper, the H{\'e}non map provides a compelling and practical alternative for evaluating RC systems.

It is also worth noting that, unless considerable computational resources are employed and judicious fine-tuning of the hyperparameters in the RC algorithm is performed \cite{Luk09, Gau21, Mak23_review, Har24}, an RC system cannot be universally applied to all benchmarking time series. This limitation arises because the temporal behavior of each time series is inherently unique \cite{Mak21_ESN} and even sophisticated algorithms require additional adjustments---such as the learning rate $\alpha$---to effectively capture the underlying dynamics. Thus, the choice of hyperparameters significantly influences the ability of the reservoir to generalize across different types of nonlinear and chaotic datasets, reinforcing the necessity of problem-specific optimization when applying RC models. 

Nevertheless, to address the challenge of problem specificity, we extend our evaluation beyond the H{\'e}non map test by demonstrating the capability of the bubble-based RC system to successfully perform classification tasks---a less conventional application for RC systems but one that represents a valuable and intriguing extension of their capabilities \cite{Dud23, Abb24}. Unlike traditional RC applications, which primarily focus on time-series prediction and forecasting, classification tasks require the system to identify patterns and assign inputs to distinct categories based on learned features.

\subsection{Computational operating regimes}
In addition to the ability of the bubble-based RC system to operate in either a periodic or chaotic regime---depending on physical system parameters and input conditions---its performance can be evaluated through various computationally defined metrics.

Typically, the first computational test involves the predictive mode, where the RC system is presented with a previously unseen data point and then predicts the next point, continuing iteratively \cite{Luk12, Mak23_EPL}. While this test is relatively simple, it helps assess the quality of the training process and prepares the RC system for more challenging tasks.

The second test used in this paper is the generative mode, also known as the free-running forecast \cite{Luk12}. In this mode, the output produced by the trained reservoir at the previous time step serves as the input for the next time step \cite{Luk12}, effectively making the reservoir a self-generator \cite{Sho23}. Operating in generative mode is a more challenging task compared with the predictive mode, but it holds greater practical significance, as generative reservoirs can be applied to a wide range of problems, including the prediction of complex, hard-to-analyze processes such as financial market behavior and climate variations \cite{Mak23_review}.
\begin{figure}[t]
    \centering
    \includegraphics[width=\linewidth]{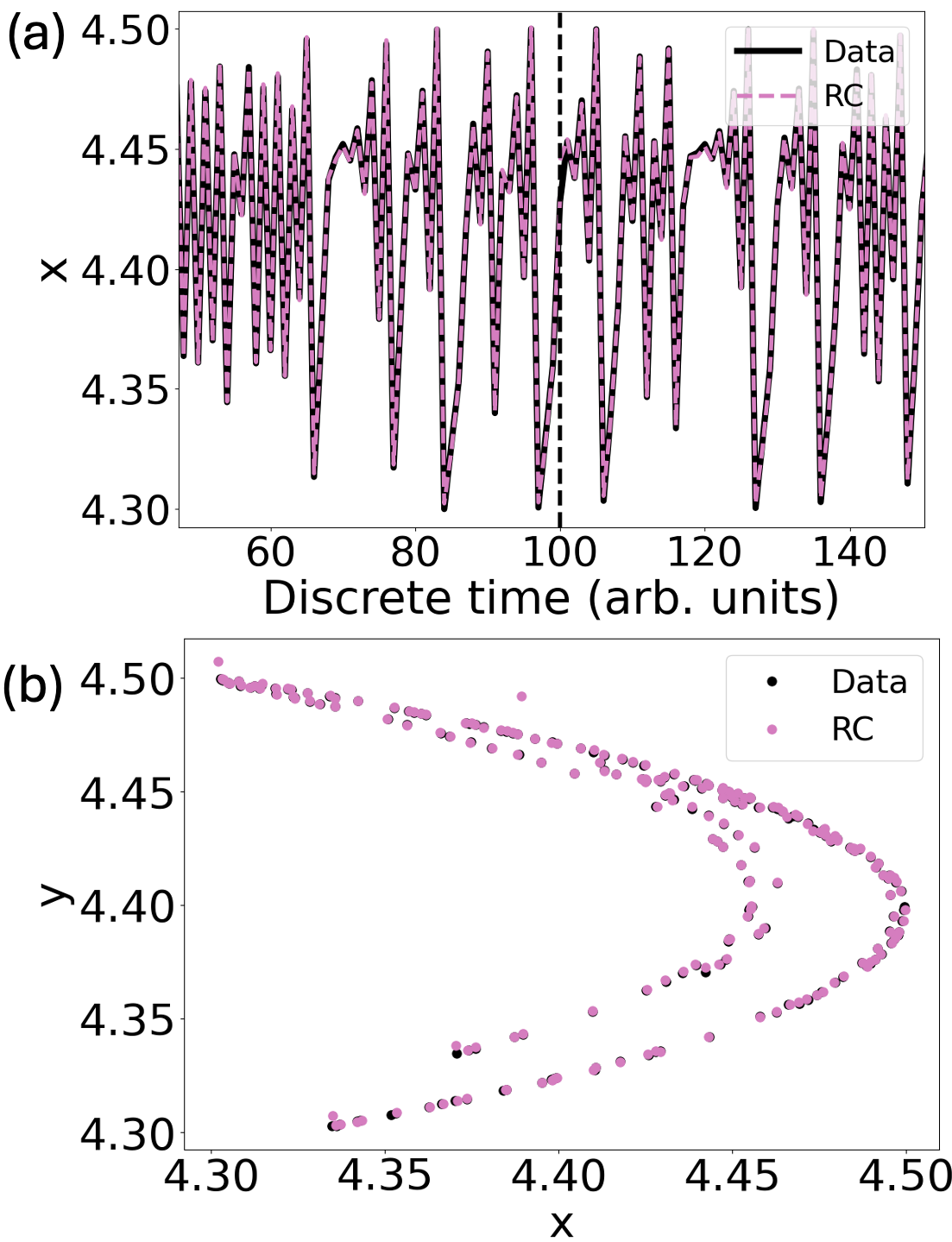} 
    \caption{Predictive mode output of the bubble-based RC systems operating in the chaotic regime. (a)~The $x$-components of the H{\'e}non map. The data points to the left and right of the vertical dashed line represent the performance of the RC system in the training and exploitation regimes, respectively. (b)~Two-dimensional representation of the predicted H{\'e}non map, with the ground truth data points shown as black dots and predicted points marked in magenta.}
    \label{fig19}
\end{figure}

\section{Results and Discussion\label{sec:Results}}
\subsection{Predictive Mode}
We begin by discussing the performance of the RC system in the predictive mode, evaluating its accuracy in forecasting the H{\'e}non map. This is quantified by computing the NMSE values in both the periodic and chaotic physical regimes of bubble oscillations. 

Figure~\ref{fig19}(a) shows a typical predictive mode output of the bubble-based RC system operating in the chaotic regime, presenting a one-dimensional plot of the $x$-components of the H{\'e}non map, with an NMSE value of approximately $8 \times 10^{-3}$. Figure~\ref{fig19}(b) presents the corresponding two-dimensional representation of the H{\'e}non map, with the ground truth data points shown as black dots and predicted points marked in magenta.

\begin{figure}[t]
    \centering
    \includegraphics[width=\linewidth]{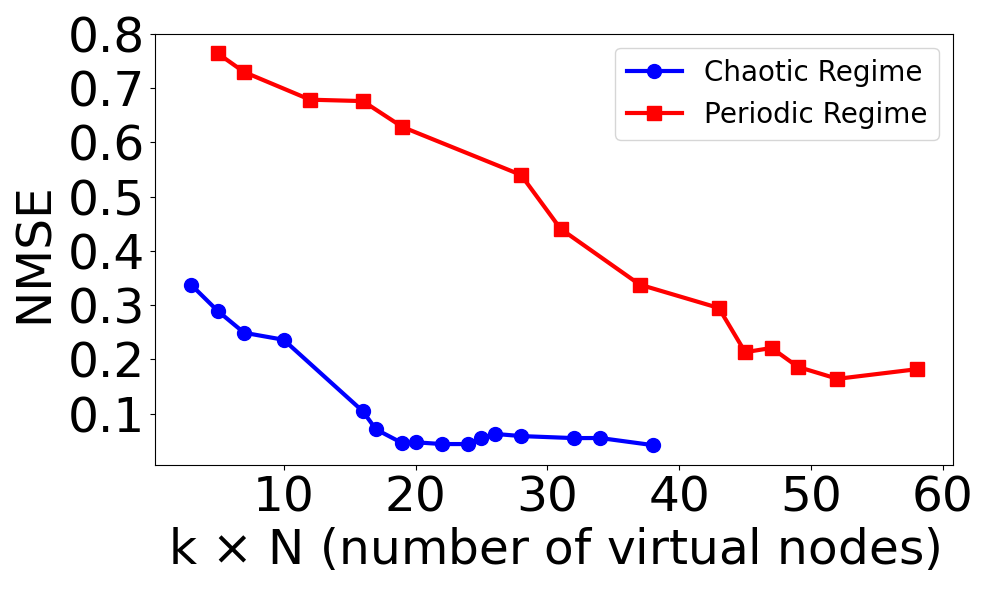}
    \caption{Predictive mode NMSE plotted as a function of the reservoir size, defined as the total number of virtual nodes $k \times N$, for both periodic and chaotic physical regimes of bubble oscillation.}
    \label{fig17}
\end{figure}

Although a visually similar result was also achieved by selecting the periodic physical oscillation regime, as shown in Fig.~\ref{fig17}, the NMSE obtained in the chaotic regime is not only lower than that in the periodic regime but also converges more quickly. Importantly, the result in Fig.~\ref{fig17} demonstrates that a minimal configuration of just $k \times N = 15$ virtual neurons is sufficient to achieve reliable RC system performance in the chaotic regime.  

Remarkably, such a low number of neurons---compared to at least 1000 in highly optimized traditional algorithmic RC systems \cite{Luk12}---was previously observed primarily in quantum-mechanical reservoirs \cite{Dud23, Abb24}. This finding confirms the ability of a single oscillating bubble to function as an efficient, classical neuromorphic computing unit. In fact, the requirement of more than 1000 neurons in a traditional algorithmic RC system implies that computations must handle large matrices that need to be stored in computer memory and processed using complex linear algebra algorithms in a loop spanning all input and output data points \cite{Luk12}. Therefore, the introduction of virtual neurons, extracted from the nonlinear dynamical behavior of the oscillating bubble using the procedure proposed in this paper, significantly reduces computational requirements compared to traditional algorithms.

\subsection{Free-Running Mode}
\begin{figure}[th]
    \centering
    \includegraphics[width=\linewidth]{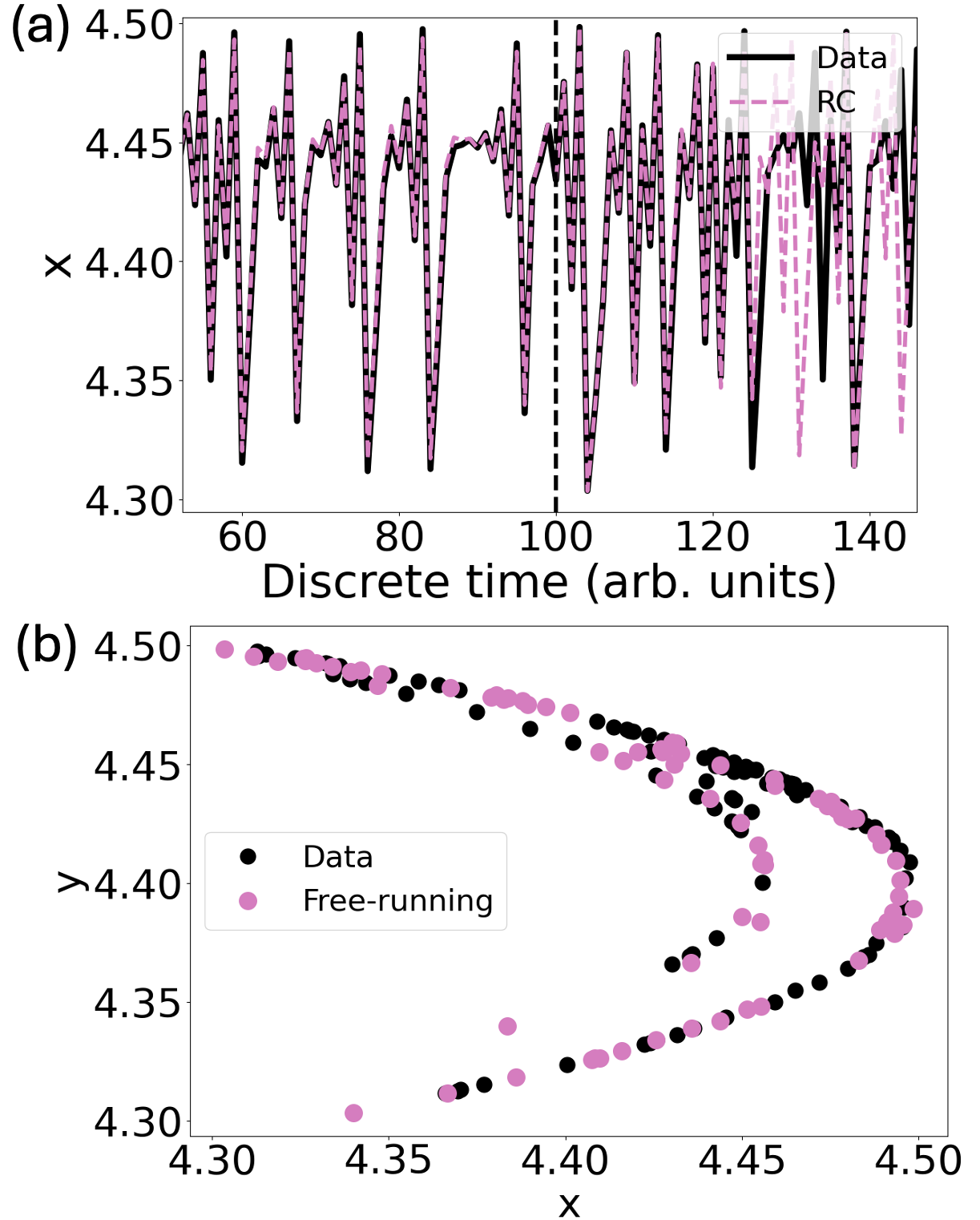}
    \caption{Free-running mode output of the bubble-based RC systems operating the chaotic regime. (a) A one-dimensional plot of the $x$-components of the H{\'e}non map. The RC system is trained on the training portion of the time series (up to the vertical dashed line) and an iterative windowing approach is used to generate predictions. (b) Two-dimensional phase space representation of H{\'e}non map, where the data points are shown as black dots and the generated points from the RC system are marked with magenta dots.}
    \label{fig18}
\end{figure}

The computational advantage of the bubble-based reservoir extends to tests conducted in the free-running mode, where the RC system autonomously generates future time series points without external input, relying solely on its internal state dynamics. We reveal that in the chaotic regime [Fig.~\ref{fig18}(a)], the bubble reservoir accurately predicts a substantial segment of the H{\'e}non time series before a noticeable deviation from the target data emerges. We can see that the predicted trajectory closely follows the ground truth within this range, indicating short-term predictive capability of the reservoir. However, beyond this range, error accumulation and sensitivity to initial conditions lead to divergence, which is a common physical characteristic of chaotic systems and a fundamental computational behavior of ML algorithms designed to predict highly nonlinear time series \cite{Luk12, Mak21_ESN, Mak23_review}.

Interestingly, we can also see that after a region of significant divergence from the ground truth the output of the RC system returns to convergence for a short period of time before diverging again. This behavior has also been reported in other physical RC systems \cite{Mak23_review}. Prolonged regions of good convergence occur when the dynamics of the reservoir temporarily aligns with the true behavior of the time series of interest. Since the memory of the reservoir is fundamentally limited \cite{Jae01}, as well as because noise and numerical instabilities amplify small deviations over time, divergence from the ground truth occurs. Nevertheless, intermittent reconvergence can occur when the dynamics of the reservoir realigns with that of the time series, particularly when the input data exhibit recurrent dynamical patterns \cite{Pat17, Pat18, Gau21}.

We reconstruct the two-dimensional H{\'e}non map using an iterative windowing computational approach. In this method, a sliding window of 20 consecutive points is used to generate predictions. In the first iteration, the RC system is trained on the original training data [the data points to the left of the vertical line in Fig.~\ref{fig18}(a)], resulting in a partially converged prediction compared to the ground truth. In subsequent iterations, the first 20 points of the training dataset are omitted but the first 20 points correctly predicted by the RC system are appended to the end of the training dataset. This process is repeated iteratively, continuously updating the dataset with newly generated points, until the two-dimensional H{\'e}non map is reconstructed [Fig.~\ref{fig18}(b)]. As shown, this approach yields feasible results using only a few iterations, requiring minimal additional computer memory and only slightly increasing the computational time needed for forecasting.

Of course, advanced algorithmic RC systems implemented on high-performance hardware \cite{Gau21, Har24} may achieve a broader range of high divergence between the forecast and ground truth. However, they require a significantly larger number of neurons and input data points to train the reservoir. In contrast, as indicated by the vertical dashed line in Fig.~\ref{fig18}(a), the bubble-based reservoir can operate effectively with just 100 training points. This capability is particularly advantageous for unconventional computing systems designed to operate onboard autonomous vehicles \cite{Abb24_review}.


\subsection{Classification task}
\begin{figure}[t]
    \centering
    \includegraphics[width=\linewidth]{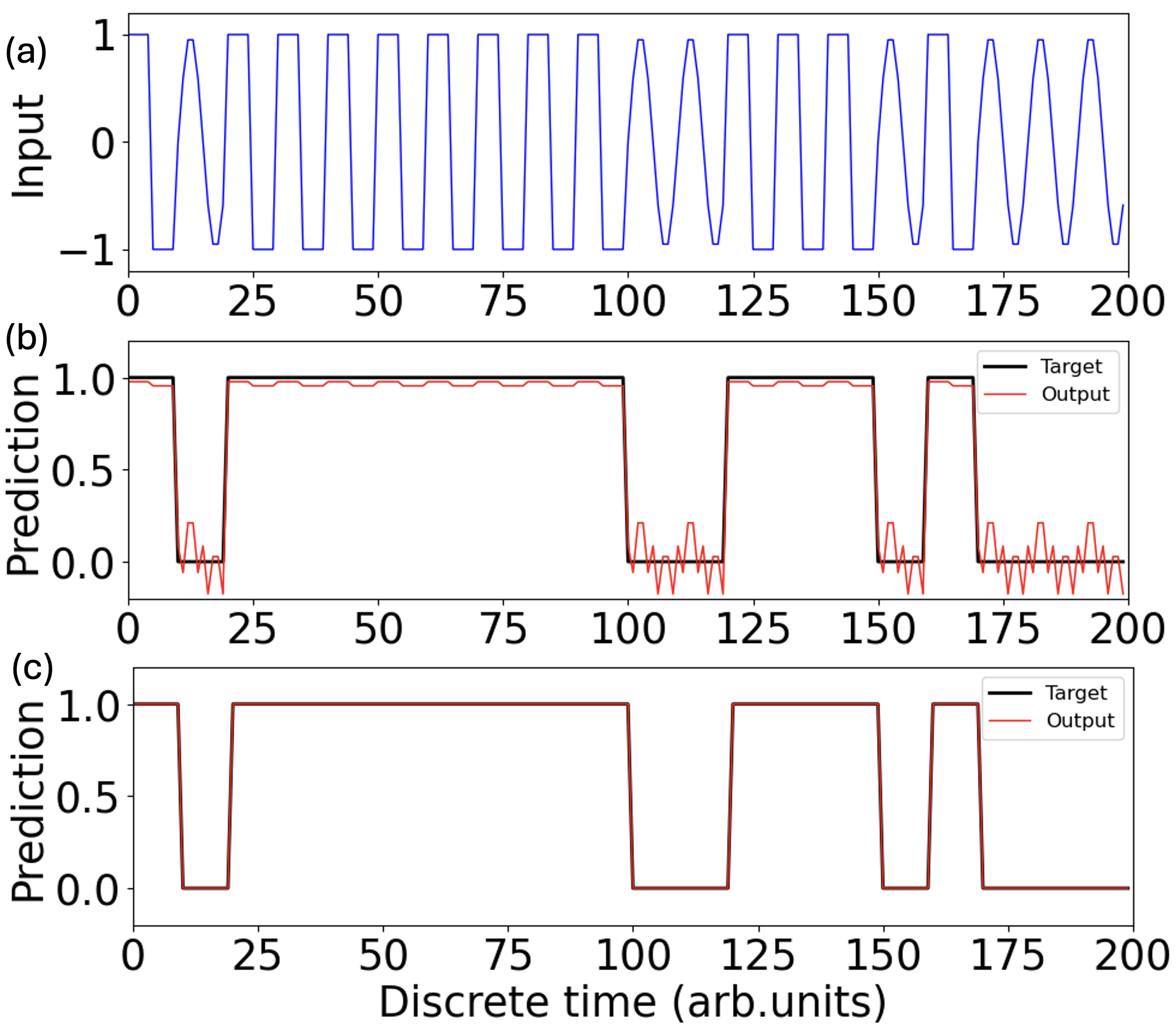}
    \caption {(a)~Classification task of sinusoidal and square waveform be the bubble-based RC equipped with (b)~6 and (c)~16 neurons, respectively.}
    \label{fig:classification}
\end{figure}

In addition to the tests on chaotic time series, we evaluate the performance of the bubble-based RC system by tasking it with a binary classification problem involving a synthetic waveform composed of randomly generated sinusoidal and square pulses [Fig.~\ref{fig:classification}(a)]. In this task, the reservoir output is expected to be 0 for square wave segments and 1 for sinusoidal segments. Correct answers of the trained RC system in this task will signify that the reservoir has both sufficient memory capacity and high nonlinearity, which are required to undertake numerous practical tasks \cite{Luk09, Dud23, Abb24}.

Figure~\ref{fig:classification}(b,~c) present the classification results produced by the bubble-based RC system equipped with 6 and 16 neurons, respectively. The solid black line represents the target output, while the red line shows the output of the reservoir. We note that the target signal is displayed on this figure only for comparison purposes and is not provided to the RC system during the exploitation stage. For the configuration with 6 neurons, the system achieved a NRMSE of $6.77 \times 10^{-2}$, while for 16 neurons the NRMSE improved to $8.94 \times 10^{-4}$, resulting in nearly perfect graphical accuracy.

\section{Conclusions}
In this work, we have presented a detailed study on the performance of a bubble-based RC system for both time series prediction and classification tasks. Our results confirm that this physical reservoir can effectively handle nonlinear and chaotic data while requiring fewer computational resources compared to traditional RC approaches.

The bubble-based RC system exhibits strong predictive performance for chaotic time series, as demonstrated by the H{\'e}non map test, while requiring only a minimal number of virtual neurons. It operates effectively in both periodic and chaotic physical regimes, with the chaotic regime providing computational advantages by capturing rich temporal dependencies. Results from the free-running mode confirm that the bubble-based reservoir can autonomously generate future time series points. Furthermore, the classification test highlights its high accuracy in pattern recognition, showcasing its potential for applications beyond conventional RC frameworks.

Future research will focus on optimizing the parameter space of the bubble-based physical system, exploring additional nonlinear datasets and investigating practical implementations in real-world scenarios such as environmental monitoring, autonomous vehicle guiding and biomedical signal processing. Furthermore, future work will explore the use of different bubble sizes to predict various time series and the implementation of multiple reservoirs to enhance the performance of bubble-based prediction tasks.

\bibliography{ref}

\end{document}